\documentclass[twocolumn,aps,prl,preprintnumbers,reprint,floatfix]{revtex4-1}
\usepackage{bbm}
\usepackage{bm}
\usepackage{float}
\usepackage{ulem}
\usepackage{amsmath}
\usepackage{amssymb}
\usepackage{empheq}
\usepackage{graphicx}
\usepackage{mathrsfs}
\usepackage{amsfonts}
\usepackage{amsthm}
\usepackage{color}
\usepackage{bigints}
\usepackage{txfonts}
\usepackage{hyperref}
\hypersetup{
     unicode=false,          
     pdftoolbar=true,        
     pdfmenubar=true,        
     pdffitwindow=false,     
     pdfstartview={FitH},    
     pdftitle={My title},    
     pdfauthor={Author},     
     pdfsubject={Subject},   
     pdfcreator={Creator},   
     pdfproducer={Producer}, 
     pdfkeywords={keyword1} {key2} {key3}, 
     pdfnewwindow=true,      
     colorlinks=false,       
     linkcolor=red,          
     citecolor=green,        
     filecolor=magenta,      
     urlcolor=cyan           
}

\providecommand{\U}[1]{\protect\rule{.1in}{.1in}}
\makeatletter
\@ifundefined{textcolor}{}
{
\definecolor{BLACK}{gray}{0}
\definecolor{WHITE}{gray}{1}
\definecolor{RED}{rgb}{1,0,0}
\definecolor{GREEN}{rgb}{0,1,0}
\definecolor{BLUE}{rgb}{0,0,1}
\definecolor{CYAN}{cmyk}{1,0,0,0}
\definecolor{MAGENTA}{cmyk}{0,1,0,0}
\definecolor{YELLOW}{cmyk}{0,0,1,0}
}
\makeatother

\begin{document}

\title{Level statistics of extended states in random non-Hermitian Hamiltonians}
\author{C. Wang}
\affiliation{School of Electronic Science and Engineering
and State Key Laboratory of Electronic Thin Films and
Integrated Devices, University of Electronic Science 
and Technology of China, Chengdu 610054, China}
\affiliation{Center for Joint Quantum Studies and Department of Physics, 
School of Science, Tianjin University, Tianjin 300350, China}
\author{X. R. Wang}
\email[Corresponding author: ]{phxwan@ust.hk}
\affiliation{Physics Department, The Hong Kong University of Science 
and Technology (HKUST), Clear Water Bay, Kowloon, Hong Kong}
\affiliation{HKUST Shenzhen Research Institute, Shenzhen 518057, China}
\date{\today}

\begin{abstract}
Absence of level repulsion between extended states in random non-Hermitian 
systems is demonstrated. As a result, the general Wigner-Dyson distributions 
of level spacing of diffusive metals in the usual Hermitian systems is replaced 
by the Poisson distribution for quasiparticle level spacing of non-Hermitian 
disordered metals in the thermodynamic limit of infinite system size. 
This is a very surprising result because Poisson statistics is universally 
true for the Anderson insulators where energy eigenstates do not overlap 
with each other so that energy levels are independent from each other.
For disordered metals where different eigenstates overlap with each 
other, one should expect different levels trying to stay away from 
each other so that the Poisson distribution should not apply there. 
Our results show that the larger non-Hermitian energy (dissipation) can 
invalidate level repulsion principle that holds dearly in quantum mechanics. 
Thus, our theory provides a unified picture for recent discovery of so 
called ``level attraction'' in various systems. 
It provides also a theoretical basis for manipulating energy levels.
\end{abstract}
\maketitle

\section{Introduction}
\label{sec1}

Open systems described by non-Hermitian Hamiltonians have drawn increasing 
attention in recent years \cite{hjcarmichael1,smalzard1,sdiehl1,bzhen1,
sbittner1,aregensburger1,sklaiman1,ceruter1,slonghi1,kgmakris1,telee1,
ychoi1,hcao1,lfeng1,hhodaei1,lchang1,tgao1,bpeng1,kesaki1,sdliang1,telee2,
dleykam1,yxu1,hmenke1,yxiong1,vmmalvarez1,cli1,zgong1,xni1,hshen1,syao1,
syao2,tmphilip1,ychen1,cwang3,nhatano,nmshnerb,vkozii1,mpapaj1,aazyuzin1} 
because of their academic interest and importance/relevance to reality. 
Unlike Hermitian Hamiltonians whose eigenenergies are real, eigenenergies of 
non-Hermitian Hamiltonians are, in general, complex numbers whose real parts 
are interpreted as quasiparticle energies and the imaginary parts are the  
inverse of quasiparticle lifetimes \cite{vkozii1,aazyuzin1,mpapaj1,hshen1}.
It is known that the level spacing distribution of random Hermitian 
Hamiltonian is a fundamental quantity that reveals the underlying physics. 
For example, level repulsion is a general principle in Hermitian quantum 
mechanics. This principle prevents two extended states from having exactly 
the same energy and leads to the famous Wigner-Dyson distribution 
$P_{\beta}(s)=C_1s^{\beta}\exp[-C_2 s^2]$ for the nearest energy level 
spacing $s$ of extended states of random Hermitian systems \cite{mlmehta1}. 
Here $\beta=1,2,4$ are respectively for the Gaussian orthogonal, unitary, 
and symplectic ensembles whose Hamiltonian matrix elements are real, 
complex and quaternion numbers, respectively. On the other hand, the 
level statistics of non-Hermitian random matrices have also attracted 
considerable attention for a long time
\cite{fhaake1,cpoli1,rhamazaki1,aftzortzakakis1,yhuang1}. 
Among more recent works \cite{rhamazaki1,aftzortzakakis1,yhuang1}, 
a non-Hermitian type of ``level repulsion'' is observed by considering 
level spacings as distances between two nearest neighbor eigenvalues 
in the complex plane.
\par

Recently, a number of experiments \cite{mharder1,klaui,skkim,cmhu} 
suggest the quasiparticles energyies $\text{Re}[E]$ can cross each 
other in non-Hermitian systems, instead of anti-crossing universally 
arising in all Hermitian Hamiltonians. This remarkable phenomenon 
is termed as level attraction \cite{byao1}. Interesting and important 
questions are how the level attraction changes the level statistics of 
the quasiparticles energyies of these systems and whether the 
new level statistics is universal.
\par
 
In this work, we study a disordered two-dimensional electron gas (2DEG) 
subjected to a perpendicular imaginary magnetic field that models the finite 
lifetime of electronic levels due to the electron-electron, or electron-phonon 
or electron-impurity interactions \cite{vkozii1,aazyuzin1,mpapaj1,hshen1}. 
It is well known that disordered Hermitian 2DEG can support extended states in 
the absence of a magnetic field only when spin-orbit interaction is present \cite{cwang}. 
In order to facilitate a metal-insulator transition, the model Hamiltonian 
contains also a Rashba/Dresselhaus or SU(2) spin-orbit coupling (SOC) that 
widely exists in 2DEGs, especially in semiconductor heterostructures.
This non-Hermitian model supports the Anderson localization transitions 
(ALTs), similar to its Hermitian counterparts \cite{cwang2}. 
Surprisingly, spacings of quasiparticle energies $\text{Re}[E]$ 
of extended states follow the Poisson distribution $P(s)=\exp[-s]$ 
in the thermodynamic limit of infinite system size no matter 
whether the system preserves time-reversal (TR) symmetry or not. 
For a finite system when the non-Hermicity energy is smaller than mean 
level spacing, quasiparticle level spacings follow the Wigner-Dyson 
distribution $P_{\beta}(s)$. 
On the other hand, in both limits, spacing distributions of 
the imaginary parts of the complex eigenenergies $\text{Im}[E]$ of the 
extended states are also universal in the sense that they do not 
depend on the models and model parameters.
\par

The paper is organized as follows. The model and numerical methods
are described in Sec.~\ref{sec2}, while the existence of ALTs 
is substantiated in Sec.~\ref{sec3}. Various results 
of level statistics are presented in Sec.~\ref{sec4}. A discussion of the
experimental relevance and a summary are given in Sec.~\ref{sec5} 
and \ref{sec6}, respectively.

\section{Model and methods}
\label{sec2}

Our model is non-interacting electrons on a square lattice subjected to 
an imaginary magnetic perpendicular field \cite{nhatano} that generates 
a non-Hermitian term $i\gamma\sigma_z$ without skin effect \cite{syao2},  
\begin{equation}
\begin{gathered}
H=\sum_{i} c^\dagger_{i}(\epsilon_{i}\sigma_0
+\eta\sigma_z+i\gamma\sigma_z)c_{i}
+\left(t\sum_{\langle ij\rangle}c^{\dagger}_{i} V_{ij}
c_{j}+h.c.\right),
\end{gathered}\label{hamiltonian}
\end{equation}
where $c^\dagger_{i}=(c^{\dagger}_{i,\uparrow},c^\dagger_{i,\downarrow})$ 
and $c_{i}$ are electron creation and annihilation operators at lattice site 
$i=(x_i,y_i)$. 
$\sigma_0$ and $\sigma_{x,y,z}$ are respectively the two-by-two 
identity matrix and Pauli matrices acting on the spin space. 
$t=1$ is used as the energy unit. 
Randomness is introduced through $\epsilon_i/t$ that randomly distributes
in $[-W/2,W/2]$ with $W$ measuring disorder strength. 
Rashba SOC \cite{cwang2} of strength $\alpha=0.1$ encoded in two-by-two matrices 
of $V_{ij}=V_x=\sigma_0+i\alpha\sigma_y$ and $V_{ij}=V_y=\sigma_0-i\alpha\sigma_x$ 
for $\langle ij\rangle$ along the $x-$ and the $y-$directions, respectively, 
is used in this study. Note that Hamiltonian~\eqref{hamiltonian} preserves 
the TR symmetry if $\eta=0$ while the TR symmetry is broken for $\eta\neq 0$.
This can easily be checked from the TR operator 
$\mathcal{T}=-i\sigma_y K$ that commutes with the Hamiltonian 
$\mathcal{T}H\mathcal{T}^{-1}= H$ for $\eta=0$ and does not commutes 
with $H$ for $\eta\neq 0$, $\mathcal{T}H \mathcal{T}^{-1}\neq H$, 
where $K$ is the complex conjugation \cite{kkawabata1}.  
\par

The eigenstates of Hamiltonian~\eqref{hamiltonian} can be either localized 
or extended, and these two groups of states form separated bands. 
This can be seen from the inverse participation ratio (IPR) of a right 
eigenstate $\psi_E$ defined as 
$p_2(E,W)=\langle\sum_{i}|\psi_E(i)|^4\rangle^{-1}$, 
where $\psi_E(i)$ is the wave function amplitude at site $i$.  
$\psi_E$ satisfies $H|\psi_E\rangle=E|\psi_E\rangle$ and 
$\langle\psi_E|\psi_E\rangle=1$. $p_2$ measures how many lattice 
sites are occupied by the wave function. 
If there exists an ALT from extended states to localized states when 
disorder strength $W$ varies for a fixed $E$, the correlation length 
$\xi$ diverges at the critical value $W_c$ as $\xi(W)\propto|W-W_c|^{-\nu}$. 
$p_2$ near $W_c$ satisfies the following one-parameter scaling function 
\cite{xrwang1,jhpixley1,cwang4}
\begin{equation}
\begin{gathered}
p_2(W)=L^D[f(L/\xi)+C/L^y].
\end{gathered}\label{scaling}
\end{equation}
Here $f(x)$ is an unknown scaling function to be determined, $C$ is 
a constant, and $y>0$ is the exponent for the irrelevant variable. 
$D$ is the fractal dimension of critical wave functions which occupy a 
subspace of dimensionality smaller than the embedded space dimension $d=2$. 
The critical exponent $\nu$, together with the fractal dimension $D$, 
characterizes the universality class of ALTs according to the 
quantum phase transition ansatz \cite{fevers1,cwang2}.
The following criteria are used to identify an ALT: 
(1) $Y_L(W)=p_2L^{-D}-CL^{-y}$ increases and decreased with $L$ for an 
extended and a localized state, respectively.
(2) Near $W_c$, $Y_L(W)$ of different system sizes $L$ collapse into two 
branches of a smooth function (one for localized states and the other for 
extended states). The implementation of the finite-size scaling 
analysis is illustrated in detail in appendix~\ref{app0}.
\par

To compute the level statistics of the real (quasiparticle energies) and 
imaginary parts of eigenenergy $E$, we diagonalize exactly the Hamiltonian 
with periodic boundary conditions in both directions to obtained all $E$'s. 
$\text{Re}[E]$ is sorted in the ascending order. The diagonalization is 
performed by using Scipy library \cite{scipy}. We consider the eigenenergies 
in a very narrow energy window for many realiztions. The ensemble-averaged
level spacing distribution for both $\text{Re}[E]$ and $\text{Im}[E]$,
denoted as $P_R(s)$ and $P_I(s)$, respectively, can be described by 
the histogram plot, where the systematic error in the histogram plots 
is eliminate to increase the accuracy \cite{cwang2}. We also exclude the Kramers 
double degeneracy when calculating $P_R(s)$ for systems with the TR symmetry.
\par

\section{Existence of ALTs}
\label{sec3}

\begin{figure*}[htbp]
\centering
  \includegraphics[width=0.9\textwidth]{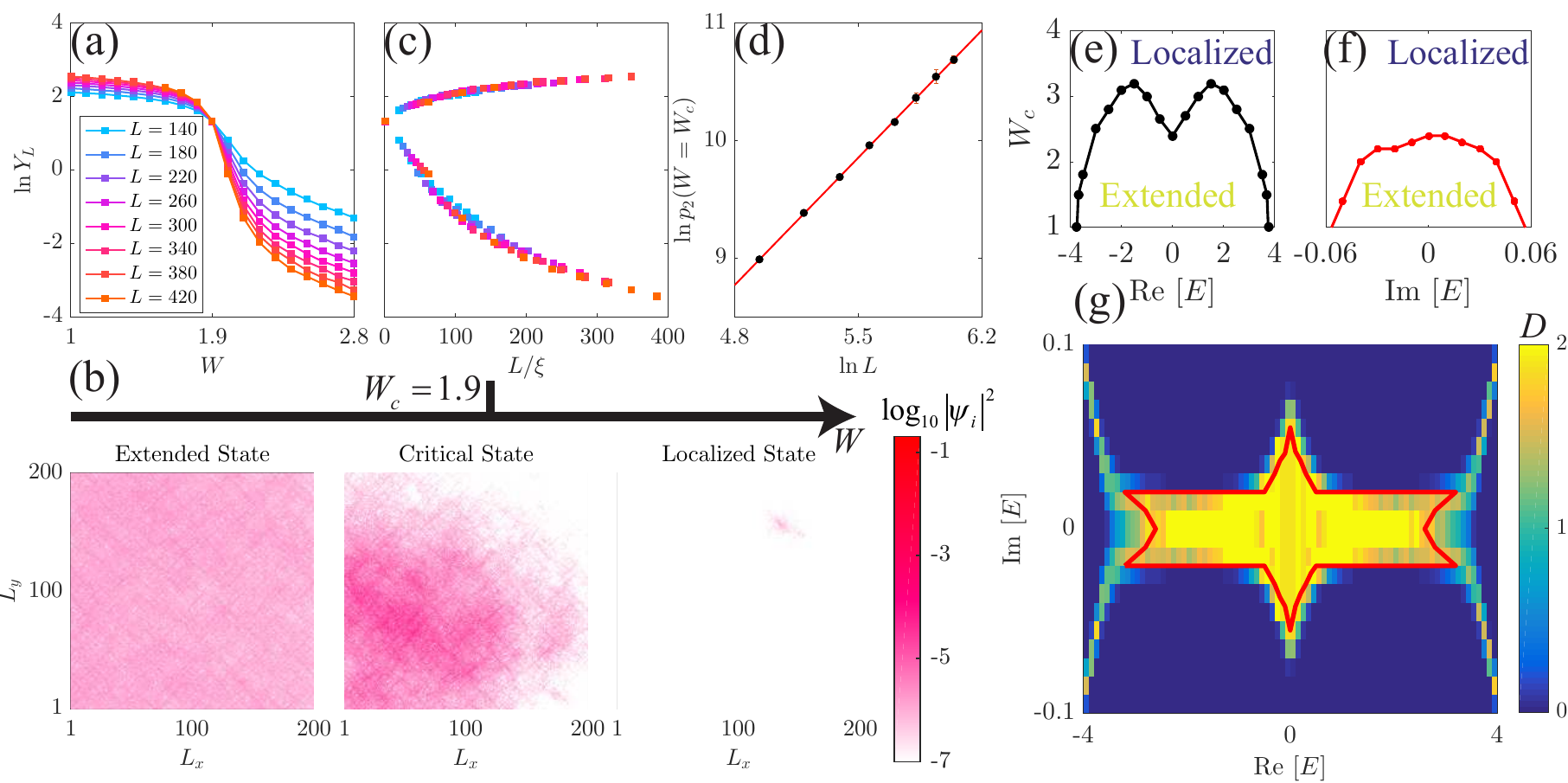}
\caption{(a) $\ln Y_L$ vs $W$ for state of $E=0$. 
(b) Spatial distributions $\log_{10}|\Psi(x_i,y_i)|^2$ 
of wave function of state of $E=0$ in a typical 
realization for $W=1$ (extended), 1.9 (critical), 
and 5 (localized). The degree of red encodes the  
probability density as indicated by the color bar. 
(c) Scaling function $\ln [f(L/\xi)]$ vs $L/\xi$. 
(d) $\ln p_2(W=W_c)$ as a function of $\ln L$. 
The solid line is a linear fit with slope $D=1.60\pm0.05$.
(e) $W_c$ vs $\text{Re}[E]$ for $\text{Im}[E]=0$. 
(f) $W_c$ vs $\text{Im}[E]$ for $\text{Re}[E]=0$.
(g) Phase diagram in the complex eigenenergy plane $E$ at 
a fixed disorder strength $W$. 
Colour encodes the fractal dimension $D$. 
The red line is the mobility boundary with $D=1.6\pm0.1$. 
Each point is averaged over 200 samples.}
\label{fig_ipr}
\end{figure*}
\par 

We first identify the ALTs from the finite-size scaling of the IPR. 
Similar to its Hermitian counterparts \cite{cwang2}, an ALT of 
system \eqref{hamiltonian} occurs at a critical disorder strength $W_c$ 
at which all curves of $\ln Y_L(W)$ as a function $W$ for a state with 
given energy $E$ and for different system size $L$ cross as shown in 
Fig.~\ref{fig_ipr}(a) for $E=0$, $\gamma=\eta=0.1$ and $L$ ranging from 
140 to 420. Indeed, data in Fig.~\ref{fig_ipr}(a) gives $W_c=1.90\pm0.02$, 
and $d \ln Y_L(W)/dL$ is positive for $W<W_c$ and negative for $W>W_c$. 
These features clearly support the occurrence of an ALT: 
The state of $E=0$ is extended for $W<W_c$ and becomes localized for $W>W_c$. 
We also plot the wave functions distribution $\log_{10}|\psi_i|^2$ for three 
disorder strengths: $W=1<W_c$, $W=W_c$, and $W=5>W_c$, as shown in 
Fig.~\ref{fig_ipr}(b) where the degree of red color encodes probability density. 
Apparently, the wave function spread uniformly over the whole lattice at a 
length scale larger than $\xi$ for $W<W_c$ while it is highly localized on the 
lattice for $W>W_c$. At $W=W_c$, the state is critical that occupies a much 
sparser space than those of $W<W_c$ and resemble a fractal object \cite{xrwang1}. 
\par
 
The chi-square fit of $p_2(W)$ with a satisfactory goodness-of-fit of $Q=0.2$ 
yields the critical exponent $\nu=0.83\pm0.06$, the fractal dimension of 
$D=1.60\pm0.05$, the irrelevant exponent of $y=0.10\pm0.03$, and $C=0.5\pm0.1$.  
Fig.~\ref{fig_ipr}(c) shows the scaling functions of $f(x)$ obtained by 
collapsing all curves in Fig. \ref{fig_ipr}(a) into a single one.  
We also plot $\ln p_2(W=W_c)$ vs $\ln L$ in Fig.~\ref{fig_ipr}(d), and the 
curve is a straight line of a slope [fractal dimension] of $D=1.60\pm 0.05$ 
\cite{xrwang1}, the same value as that from the scaling function analysis. 
Interestingly, it agrees with an analytical result obtained from the 
non-Hermitian XY model \cite{telee3}. 
\par

The important feature or the fingerprint of a quantum phase transition is the 
universality concept. It says that critical exponents such as correlation 
length exponent $\nu$ and fractal dimension $D$ do not depend on model parameters.
We carried out more calculations of IPR to show that $\nu$ and $D$ for 
the case without TR symmetry ($\eta=0.1$) do not depend (within numerical 
errors) on the strength of Rashba SOC $\alpha$, the complex eigenenergy $E$, 
and the form of disorders for $\gamma=0.1$. 
The results are summarized in Table~\ref{tab1}.
\par

\begin{table}
\caption{\label{tab1} Critical exponent $\nu$, fractal dimension 
$D$ of wave functions at the ALT, and the goodness-of-fit $Q$ for
different model parameters (Rashba SOC strength $\alpha$ and eigenenergy 
$E$) at a fixed non-Hermicity energy $\gamma=0.1$.
We consider two different types of disorders: (i) Independent uniform  
distribution (as those in the main text) of $\epsilon_i$ in the window 
of $[-W/2,W/2]$; (ii) Independent Gaussian distribution (used in 
Ref.~\cite{wwchen1}) of $\epsilon_i$ with zero mean and the variance of $W^2$.}
\begin{ruledtabular}
\begin{tabular}{llll}
& $\nu$ & $D$ & $Q$ \\
\hline
Uniform distribution & & & \\
\hline
$E=0.0,\alpha=0.2$ & $0.80\pm0.05$ & $1.65\pm0.03$ & 0.1  \\
$E=0.0,\alpha=0.3$ & $0.7\pm0.1$   & $1.7\pm 0.1$  & 0.05 \\
$E=0.01i,\alpha=0.1$ & $0.6\pm0.2$   & $1.6\pm 0.1$  & 0.08 \\
$E=0.1+0.01i,\alpha=0.1$ & $0.85\pm0.09$  & $1.63\pm 0.8$  & 0.1 \\
\hline
Gaussian distribution & & & \\
\hline
$E=0.0,\alpha=0.1$ & $0.8\pm0.1$  & $1.6\pm 0.2$  & 0.04 \\
\end{tabular}
\end{ruledtabular}
\end{table}

Figures~\ref{fig_ipr}(e) and (f) show how the critical disorder $W_c$ 
changes with the complex energy $E$: $W_c$ varies with $\text{Re}[E]$ 
for $\text{Im}[E]=0$ (e) and with $\text{Im}[E]$ for $\text{Re}[E]=0$ (f). 
All states are localized for $|\text{Re}[E]|>4$, and one needs the 
largest disorder strength (maximal $W_c$) to localize states around 
$\text{Re}[E]=\pm 1.6$. Different from its $\text{Re}[E]$-dependence, 
$W_c$ is monotonic in $|\text{Im}[E]|$.
\par

The boundary that separates the extended states from the localized states is 
a closed curve in the complex energy plane as shown in Fig.~\ref{fig_ipr}(g) 
obtained from extensive numerical calculations of the IPR for different 
$E$ and system sizes $L$ (ranging from $L=160$ to $L=320$) at $W=2$. 
The wave functions at the mobility boundary (the red line in 
Fig.~\ref{fig_ipr}(g)) are fractals with the same fractal dimension $D=1.6$. 
\par

\section{Level statistics}
\label{sec4}

\begin{figure}[htbp]
\centering
\includegraphics[width=0.48\textwidth]{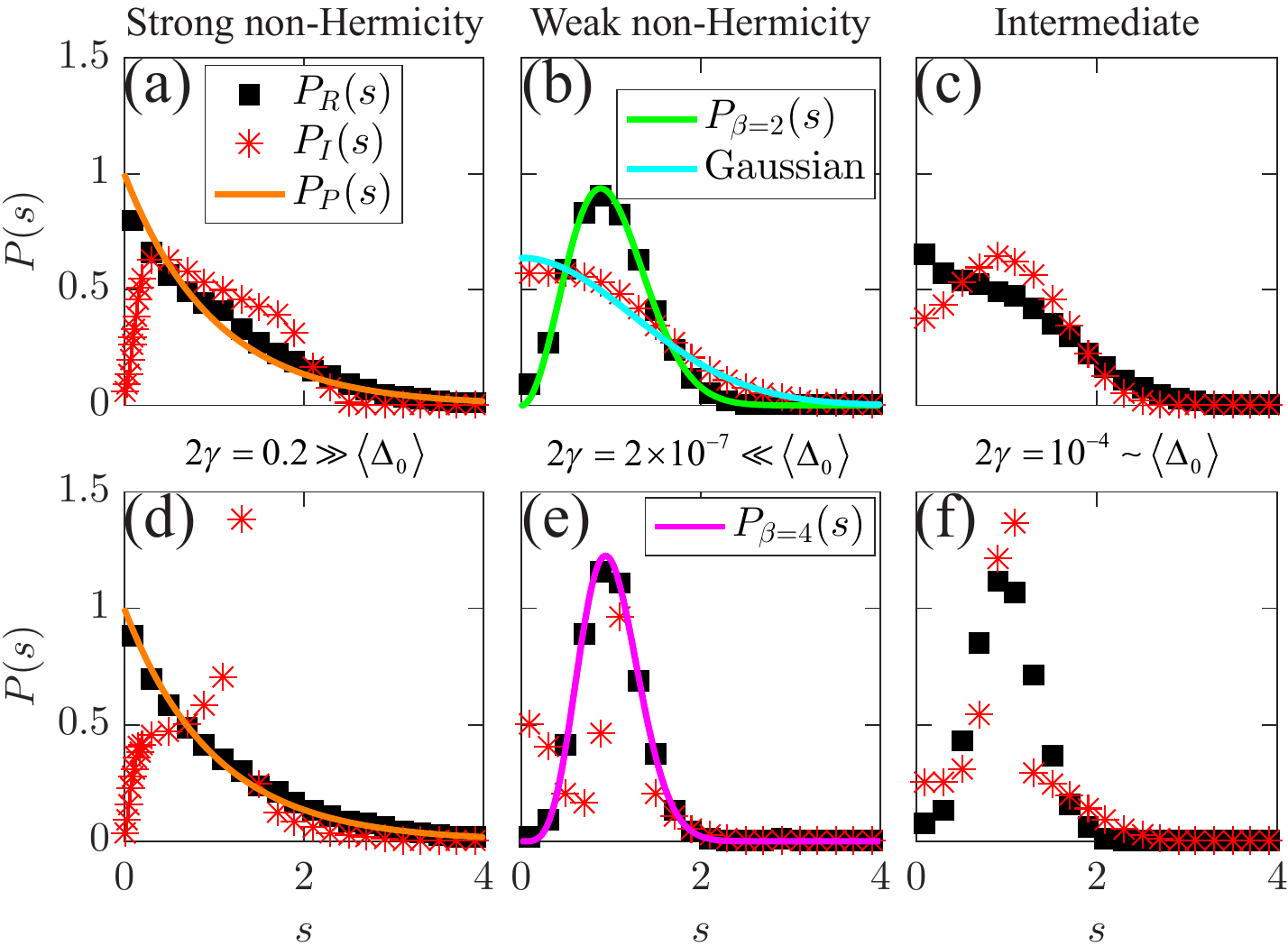}
\caption{$P_R(s)$ (the cyan squares) and $P_I(s)$ (the purple cross) in the 
cases without TR symmetry for $\eta=0.1$ (a,b,c) and with TR symmetry for 
$\eta=0$ (d,e,f) within $|E|<0.01$ for $W=1$, $L=160$, and $\gamma=0.1$ 
(a,d), $\gamma=1\times 10^{-7}$ (b,e), and $\gamma=5\times 10^{-5}$ (c,f). 
The black solid lines in (a) and (d) are $P_{\text{P}}(s)$. 
The red and the orange solid lines in (b) and (e) are $P_{\beta=2}(s)$ and 
$P_{\beta=4}(s)$, respectively. 
The green solid line in (b) is the Gaussian function.
}
\label{fig_ps}
\end{figure}

After establishing the ALTs for Hamiltonian~\eqref{hamiltonian}, we are 
now in the position to discuss the level statistics of the extended states.
Figures~\ref{fig_ps}(a) and \ref{fig_ps}(d) are $P_R(s)$ (the cyan squares) 
and $P_I(s)$ (the purple cross) for systems without TR symmetry for $\eta=0.1$ 
(a) and with TR symmetry for $\eta=0$ (b) within $|E|<0.01$ for $L=160$, $W=1$, 
and $\gamma=0.1$, where all states are extended (see Fig.~\ref{fig_ipr}). 
Surprisingly, the level-spacing distribution of $\text{Re}[E]$ is well described 
by the Poisson function $P_{\text{P}}(s)$ no matter with or without the TR symmetry, 
instead of the Wigner-Dyson distributions of $P_{\beta=2}(s)$ or $P_{\beta=4}(s)$ 
that would be the case for an Hermitian Hamiltonian when $\gamma=0$. 
This is surprising because the Poisson distribution is not normally for 
extended states, but for the localized states whose eigenenergies 
distribute independently and randomly in certain energy ranges.  
Similarly, $P_I(s)$ is universally described by an unknown function 
in the sense that it does not depend on models with different forms 
of SOCs, disorders, and dimensionality, see Appendix~\ref{app1}).
This unknown function shows a  ``level repulsion'', i.e., $P_I(s=0)=0$. 
\par 

However, for very small non-Hermicity of $\gamma=10^{-7}$ and the same 
$W=1$ and $L=160$, $P_R(s)$, obtained from those extended states within 
the window of $|E|<0.01$, follows perfectly with the Wigner-Dyson 
distributions of $P_{\beta=2}(s)$ and $P_{\beta=4}(s)$ as shown in 
Figs.~\ref{fig_ps}(b) and \ref{fig_ps}(e), respectively for the cases 
without and with TR symmetry. At the same time, $P_I(s)$ is universally 
described by the Gaussian function for $\eta\neq 0$ or by an unknown 
function with a universal non-zero constant $P_I(s=0)$, or 
non-level-repulsion, in the sense that the distribution are
model-independent, see Appendix~\ref{app1}. 
For the intermediate non-Hermicity energy of $2\gamma=10^{-4}$, some 
parameter-dependent distributions of $P_R(s)$ and $P_I(s)$ are seen, as 
shown in Fig.~\ref{fig_ps}(c) for $\eta=0.1$ and \ref{fig_ps}(f) for $\eta=0$.
\par 

To obtain the insight of the dramatical change in level statistics from 
the Wigner-Dyson distribution of $\gamma=0$ to the Poisson distribution 
of non-zero $\gamma$, we follow the wisdom of Wigner by considering the 
two-by-two non-Hermitian random matrix \cite{mlmehta1}
\begin{equation}
\begin{gathered}
\mathcal{H}=\begin{bmatrix}
\epsilon_1+\epsilon_2 & h_{12} \\
h_{21} & \epsilon_1-\epsilon_2
\end{bmatrix}+i\gamma\sigma_z.
\end{gathered}\label{2_by_2}
\end{equation} 
$\epsilon_{1,2}$ and $h_{12}$ are independent random variants of Gaussian 
distribution of zero mean and variance $\sigma^2$, i.e., 
$f(x,\sigma)\sim\exp[-x^2/\sigma^2]$. $\gamma$ is of the non-Hermicity energy. 
Hamiltonian~\eqref{2_by_2} breaks both spin-rotation symmetry and TR symmetry. 
The difference of the two eigenenergies (level spacing) is  
\begin{equation}
\begin{gathered}
\Delta=\sqrt{\Delta^2_0-4\gamma^2+i8\gamma\epsilon_2},
\end{gathered}\label{spacing}
\end{equation}
with $\Delta_0=2\sqrt{\epsilon^2_2+|h_{12}|^2}$ being the mean 
level spacing of the Hermitian part of Hamiltonian~\eqref{2_by_2}.
If $\gamma=0$, the eigenenergies are real, and its level spacing distribution 
is $P(s)=\int\delta(s-\Delta_0)\exp[-(\epsilon_2^2+|h_{12}|^2)/\sigma^2]
d\epsilon_2 d^\beta h_{12}$, where $\Delta_0=\sqrt{\epsilon_2^2+|h_{12}|^2}$, 
$|h_{12}|^2=\xi_1^2; \ \ \xi_1^2+\xi_2^2; \ \ \xi_1^2+\xi_2^2+\xi_3^2+
\xi_4^2$ for the Gaussian orthogonal ensemble ($\beta=1$, real matrix elements), 
the Gaussian unitary ensemble ($\beta=2$, complex matrix elements), and the 
Gaussian symplectic ensemble ($\beta=4$, quaternion matrix elements) respectively. 
Here $\xi_i$ ($i=1,2,3,4$) are real. Thus, $P(s)=C_1s^\beta \exp[-C_2s^2]$ is 
exactly the well-known Wigner-Dyson distribution. 
The prefactor is proportional to the area of equal-$\Delta_0$ hyper-surface in 
the $\epsilon_2-\vec\xi$ space. If $\gamma=0$ in the current case, level spacing 
$\Delta=\Delta_0$ is non-negative. Any coupling (non-zero $\xi_1$ and $\xi_2$) 
tends to push two levels apart. The probability of having zero level spacing is 
the probability to have $\epsilon_2=\xi_1=\xi_2=0$, which is vanishingly 
small and gives rise to the Wigner-Dyson distributions.
However, if $|\gamma|$ is of the order of $\Delta_0$, 
the real part of $\Delta$ is possible to be negative, zero, and positive. 
In this case, two levels can freely cross each other, and are, 
in principle, independent from each other. This is our understanding of 
why $P_R(s)$ follows the Poisson function (see derivation later).
\par 

Above poor-man's analysis reveals two relevant energy scales for the 
level statistics: The mean level spacing $\langle\Delta_0\rangle$ of 
the Hermitian part of the model and the non-Hermicity energy $2\gamma$.
We expect three different regimes.
(i) {\it Strong non-Hermicity limit} $2\gamma\gg \langle\Delta_0\rangle$: 
Level repulsion is invalid, and two quasiparticle levels can freely cross 
each other such that the quasiparticle level spacing distribution follow the 
universal Poisson function that is for independent random level distribution. 
The spacings of the imaginary part of the complex eigenenergies follow an 
unknown universal distribution function. 
(ii) {\it Weak non-Hermicity limit} $2\gamma\ll \langle\Delta_0\rangle$: 
The non-Hermicity energy is much smaller than the average level 
spacings between two Hermitian modes. Therefore, the non-Hermicity 
is not enough to induce level crossing so that quasiparticle level 
spacing of extended states follows still the Wigner-Dyson statistics. 
(iii) {Intermediate non-Hermicity:} The level spacings follow some
non-universal distributions that are sensitive to the details of a model. 
This explains well the changes of level statistics when the ratio of 
non-Hermicity energy to $\langle\Delta_0\rangle$ is tuned by fixing 
lattice size $L$ and varying $\gamma$.

\begin{figure}[htbp]
\includegraphics[width=0.48\textwidth]{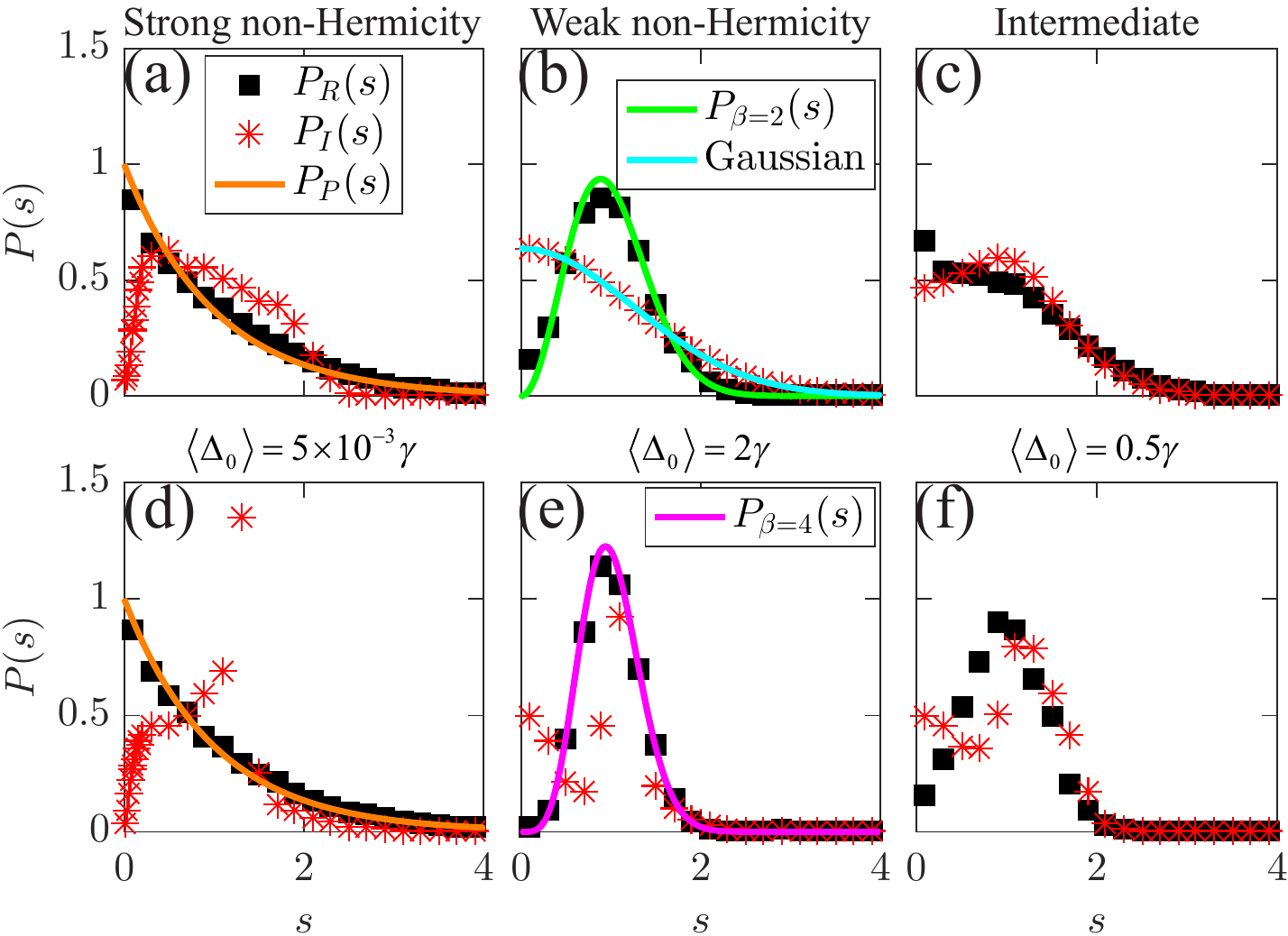}
\caption{$P_R(s)$ (cyan squares) and $P_I(s)$ (purple cross)
within $|E|<0.01$ in the cases of TR-broken ((a,b,c) for $\eta=0.1$) 
and TR-preservation ((d,e,f) for $\eta=0$) for $\gamma=10^{-2}$, 
$W=1$, and $L=200$ (a,d), $L=10$ (b,e), and $L=20$ (c,f). 
The black solid lines in (a) and (d) are $P_{\text{P}}(s)$. 
The red and the orange solid lines in (b) and (e) are $P_{\beta=2}(s)$ 
and $P_{\beta=4}(s)$, respectively. 
The green solid line in (b) is the Gaussian function. } 
\label{fig3}
\end{figure}

We further verify above picture by noticing that the ratio of 
non-Hermicity energy to $\langle\Delta_0\rangle$ can also be tuned by 
fixing $\gamma$ and varying lattice size $L$ because the mean level 
spacing is inversely proportional to the number of lattice sites as 
$\langle\Delta_0\rangle\simeq 0.22 (W+8)/L^2$, see Appendix~\ref{app2}
for the clarification. We compute $P_R(s)$ and $P_I(s)$ in the energy 
range of $|E|<0.01$ for the cases with and without TR symmetry and 
for $W=1$, $\gamma=10^{-2}$ and three different system sizes: 
$L=200$ ($\langle\Delta_0\rangle=5\times 10^{-3}\gamma$),
$L=20$ ($\langle\Delta_0\rangle=0.5\gamma$), and
$L=10$ ($\langle\Delta_0\rangle=2\gamma$).
The results are plotted in Fig.~\ref{fig3} for the cases with (a,b,c) 
and without (d,e,f) TR symmetry. 
Similar to the results for the cases of fixing $L$ and varying $\gamma$ 
above, $P_R(s)$ follows either the Poisson or Wigner-Dyson distribution 
while $P_I(s)$ follows either an unknown universal or the Gaussian 
distribution when lattice size are respectively of $L=200$ and $L=10$).
It should be noted that the system is always in the strong non-Hermicity 
limit at fixed $\gamma\neq 0$ and in the thermodynamic limit of 
$L\rightarrow\infty$ so that the quasiparticle energy level spacing 
distribution is Poissonian. All our results show that analysis based on 
the random matrix~\eqref{2_by_2} can explain the results shown in 
Figs.~\ref{fig_ps} and \ref{fig3} for strong and weak non-Hermicity limits.  
\par

Before ending this section, we would like to point out that the Poisson 
level statistics is universal for all systems without level repulsion, 
i.e. level cross each other independently as what was recently observed 
in non-Hermitian systems \cite{mharder1,klaui,skkim,cmhu}. 
If levels can freely cross, then the probability to find a nearest 
neighbouring level located within $[s, s+\delta s]$ is the product of the 
probability of no level within $[0, s]$ with the probability of the level 
falling in $[s, s+\delta s]$, i.e. $P(s)\delta s =(1-\int_0^s P(s')ds')
\delta s/\Delta$, where $\Delta$ is the mean level spacing. 
Thus, $P(s)$ satisfies differential equation of $\frac {dP}{ds} =-P/\Delta$ 
whose solution is just the Poisson function. When $\Delta $ is used as the 
unit of level spacing, $P(s)$ is exactly what we found in this paper.
\par

\section{Discussion}
\label{sec5}

There are some very recent studies of the level statistics of
non-Hermitian systems. Hamazaki {\it et al} have also
observed the Poisson distribution of $P_R(s)$ in a non-Hermitian 
many-body Hamiltonian with the TR symmetry \cite{rhamazaki1}. 
On the other hand, a non-Hermitian type level repulsion is 
witnessed by studying the distribution of spacings of two 
nearest neighbor eigenvalues in the complex energy 
space \cite{aftzortzakakis1,yhuang1}. 
Moreover, a new universal level statistics at metal-insulator 
transition is conjectured. These papers 
indeed studied the similar issue, but did not obtain the central 
results in this work, i.e., the universal Poisson distribution of 
$P_R(s)$ and $P_I(s)$ in both strong and weak non-Hermicity limits.
Obviously, our results offer a way to manipulate energy levels. 
For example, one may change the relative position of two levels 
by active level repulsion or level crossing through controlling 
the strength of non-Hermicity, a concept of damping engineering. 

Pertaining to the relevance of the reality, the Hermitian part of 
Hamiltonian ~\eqref{hamiltonian} is usually used to describe 2DEGs 
of semiconductors heterostructures with Rashba SOCs \cite{tando1}. 
The non-Hermicity term $i\gamma\sigma_z$ with an additional
non-Hermitian on-site energy $-i\gamma_0\sigma_0$ 
($\gamma_0>|\gamma|$) can arise from the spin dependent lifetimes 
due to the omnipresent electron-electron, electron-impurity, and 
electron-phonon interactions \cite{vkozii1,hshen1,aazyuzin1,mpapaj1}, 
if the semiconductors heterostructures are magnetic. In principle, 
the additional term $-i\gamma_0\sigma_0$ does not change the
level statistics discussed here, see the proof in Appendix~\ref{app3}. 
Furthermore, Rashba SOCs can emerge in cold-atomic \cite{amanchon,gorso1}, 
photonics \cite{llu1}, magnonic \cite{yonose,xswang}, and skyrmionic 
systems \cite{hyang1,zxli}. All these systems can be described by 
very similar non-Hermitian Hamiltonians due to the inevitable 
gain/loss in open systems.    

\section{Conclusion}
\label{sec6}

In conclusion, 2DEGs subjected to an imaginary magnetic field, random 
on-site energies, and SOCs undergo an ALT at a finite disorder $W_c$. 
Near $W_c$, correlation lengths diverge as $\xi(W)\propto|W-W_c|^{-\nu}$ 
with $\nu=0.83\pm0.05$. A mobility boundary separating the extended 
from the localized states exists in the complex energy plane. 
In the thermodynamic limit of infinity system size, the quasiparticle 
level spacing $P_R(s)$ in the metallic phase is universally described by 
the Poisson distribution no matter whether the system has the time-reversal 
symmetry or not, while the spacing of the imaginary part of the complex 
eigenenergies $P_I(s)$ is also universal, exhibits ``level repulsion'', 
and is sensitive to the TR symmetry. For a finite system when the 
non-Hermicity energy $\gamma$ is smaller than the mean level spacing, 
$P_R(s)$ can be described by the Wigner-Dyson distribution $P_{\beta}(s)$ 
and $P_I(s)$ is universal with a universal non-zero constant.
\par  

\section{Acknowledgments}

This work is supported by the National Natural Science Foundation of China 
(Grants No.~11774296, 11704061, and 11974296) and Hong Kong RGC (Grants 
No.~16301518, 16301619 and 16300117). C.W. is supported by UESTC and the China 
Postdoctoral Science Foundation (Grants No.~2017M610595 and 2017T100684).
C.W. also acknowledges the kindly help from Xiansi Wang and Jie Lu.
\par

\appendix

\section{Finite-size scaling analysis}
\label{app0}

To extract the fractal dimension, the critical disorder, and the critical
exponent $\nu$ at the quantum phase transitions defined in the scaling 
function Eq.~\eqref{scaling} with $\xi=\xi_0|W-W_c|^{-\nu}$, i.e.,
\begin{equation}
\begin{gathered}
p_2(L,W)=L^D [f(L|W-W_c|^{\nu}/\xi_0)+CL^{-y}],
\end{gathered}\label{app0_1}
\end{equation} 
we perform a Taylor expansion of the scaling function $f(x)$ up to
the third order in $|W-W_c|^{\nu}$ near $W=W_c$,
\begin{widetext}
\begin{equation}
\begin{gathered}
f(L|W-W_c|^{\nu}/\xi_0)=
F_0+F_1 L|W-W_c|^{\nu}/\xi_0+F_2(L|W-W_c|^{\nu}/\xi_0)^2+
F_3(L|W-W_c|^{\nu}/\xi_0)^3\\
=F_0+\tilde{F}_1L|W-W_c|^{\nu}+\tilde{F}_2(L|W-W_c|^{\nu})^2
+\tilde{F}_3(L|W-W_c|^{\nu})^3,
\end{gathered}\label{app0_2}
\end{equation} 
\end{widetext}
with $D,C,y,\nu,W_c,F_0,\tilde{F}_1,\tilde{F}_2,\tilde{F}_3$ 
being fitting parameters. Then we adjust those parameters to 
minimize the chi square
\begin{equation}
\begin{gathered}
\chi^2=\sum^{N_w}_{i}\sum^{N_l}_j\left(
\dfrac{p_2(W_i,L_j)-L^D_j[f(L_j|W_i-W_c|^{\nu}/\xi_0)+CL^{-y}_j]}{\sigma_{ij}}
\right)
\end{gathered}\label{app0_3}
\end{equation}
following the approach illustrated in the appendix of Ref.~\cite{wwchen1},
where $N_w$ and $N_e$ are the number of $W$ and $L$, respectively. 
The fitting process yields the critical disorder $W_c$,
the fractal dimension $D$, and the critical exponent $\nu$.
After determining the minimal chi square, we calculate the 
goodness-of-fit $Q$ by the standard algorithm suggested in
Ref.~\cite{numerical}, which measures how well our numerical data of 
$p_2$ fit to the model of Eq.~\eqref{app0_1}. 
Take data in Fig.~\ref{fig_ipr}(a) as examples: Following the 
above process, we obtain $Q=0.2>10^{-3}$, a satisfactory number 
that says the fit acceptable.
\par

\section{Model-independence of level statistics}
\label{app1}

To demonstrate that $P_R(s)$ and $P_I(s)$ are universal in the strong 
and weak non-Hermicity limits, 
we study level statistics of extended states
for other random non-Hermitian models with different
forms of SOCs, disorders, and dimensionality.
\par

Firstly, we study Hamiltonian ~\eqref{hamiltonian} with 
different forms of SOCs. The first one is the random 
SU(2) model subjected to an imaginary perpendicular 
magnetic field $(0,0,i\gamma)$ \cite{cwang}, 
\begin{equation}
\begin{gathered}
H=\sum_{i} c^\dagger_{i}(\epsilon_{i}
\sigma_0+\eta\sigma_z+i\gamma
\sigma_z)c_{i}+\left(t\sum_{\langle ij\rangle}c^{\dagger}_{i} 
V_{ij}c_{j}
+h.c.\right),
\end{gathered}\label{app1_1}
\end{equation}
with 
\begin{equation}
\begin{gathered}
V_{ij}=
\begin{bmatrix}
e^{-i\alpha_{ij}}\cos(\beta_{ij}/2) & 
e^{-i\gamma_{ij}}\sin(\beta_{ij}/2) \\
e^{i\gamma_{ij}}\sin(\beta_{ij}/2)  &
e^{i\alpha_{ij}}\cos(\beta_{ij}/2).
\end{bmatrix}.
\end{gathered}\label{app1_2}
\end{equation}
Here $\alpha_{ij}$ and $\gamma_{ij}$ distribute
randomly and uniformly in the range of $[0,2\pi)$.
$\sin(\beta_{ij}/2)$ distributes uniformly in $[0,1)$. 
The second model is to replace the Rashba SOC in 
model~\ref{hamiltonian} by the Dresselhaus 
SOC \cite{gdresselhaus}, where the matrices 
$V_{ij}$ are parametrized as $V_x$ and $V_y$ for 
the $x-$ and the $y-$direction hopping, respectively, 
\begin{equation}
\begin{gathered}
V_{x}=\sigma_0+i\zeta\sigma_x,
V_{y}=\sigma_0-i\zeta\sigma_y.
\end{gathered}\label{app1_3}
\end{equation}
Here the constant $\zeta$ measures the strength of the 
Dresselhaus SOC.
\par

\begin{figure}[htbp]
\centering
\includegraphics[width=0.45\textwidth]{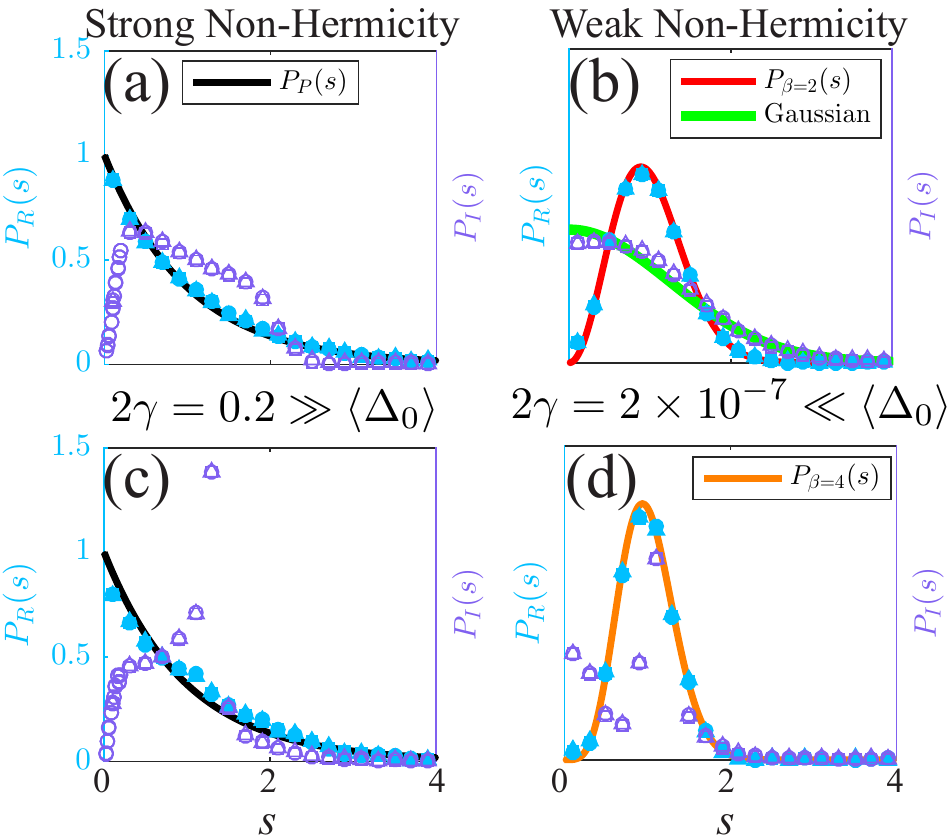}
\caption{$P_R(s)$ (the filled symbols) and $P_I(s)$ (the open symbols)
within $|E|<0.01$ in the cases without TR symmetry ((a,b) for $\eta=0.1$) 
and with TR symmetry ((c,d) for $\eta=0$) for $W=1$, $L=160$, 
and $\gamma=0.1$ (a,c), $\gamma=1\times 10^{-7}$ (b,d).
The black solid lines in (a) and (c) are $P_{\text{P}}(s)$. 
The red and the orange solid lines in (b) and (d) 
are $P_{\beta=2}(s)$ and $P_{\beta=4}(s)$, respectively. 
The green solid line in (b) is the Gaussian function.
The squares, triangles and circles are respectively 
for the Rashba SOC, the Dresselhaus SOC, and the SU(2) SOC. 
}
\label{fig4}
\end{figure}

The case without TR symmetry ($\eta= 0.1$) and the case with 
TR symmetry ($\eta= 0$) are investigated. $P_R(s)$ and $P_I(s)$ within 
the energy window of $|E|<0.01$ for $W=1$, $L=160$, 
and $\gamma=0.1$ (strong non-Hermicity limit) or 
$\gamma=10^{-7}$ (weak non-Hermicity limit) for all 
three models are plotted in Fig.~\ref{fig4}. It is 
clear that all three models (Rashba, Dresselhaus and random 
SU(2) SOCs) give identical $P_R(s)$ and $P_I(s)$. Within  
the symbol size, we cannot see any difference in 
both $P_R(s)$ and $P_I(s)$ for all three models. 
Thus, these results provide strong evidence that the new 
distributions are independent of the forms of SOCs.
\par

\begin{figure}[htbp]
\centering
  \includegraphics[width=0.45\textwidth]{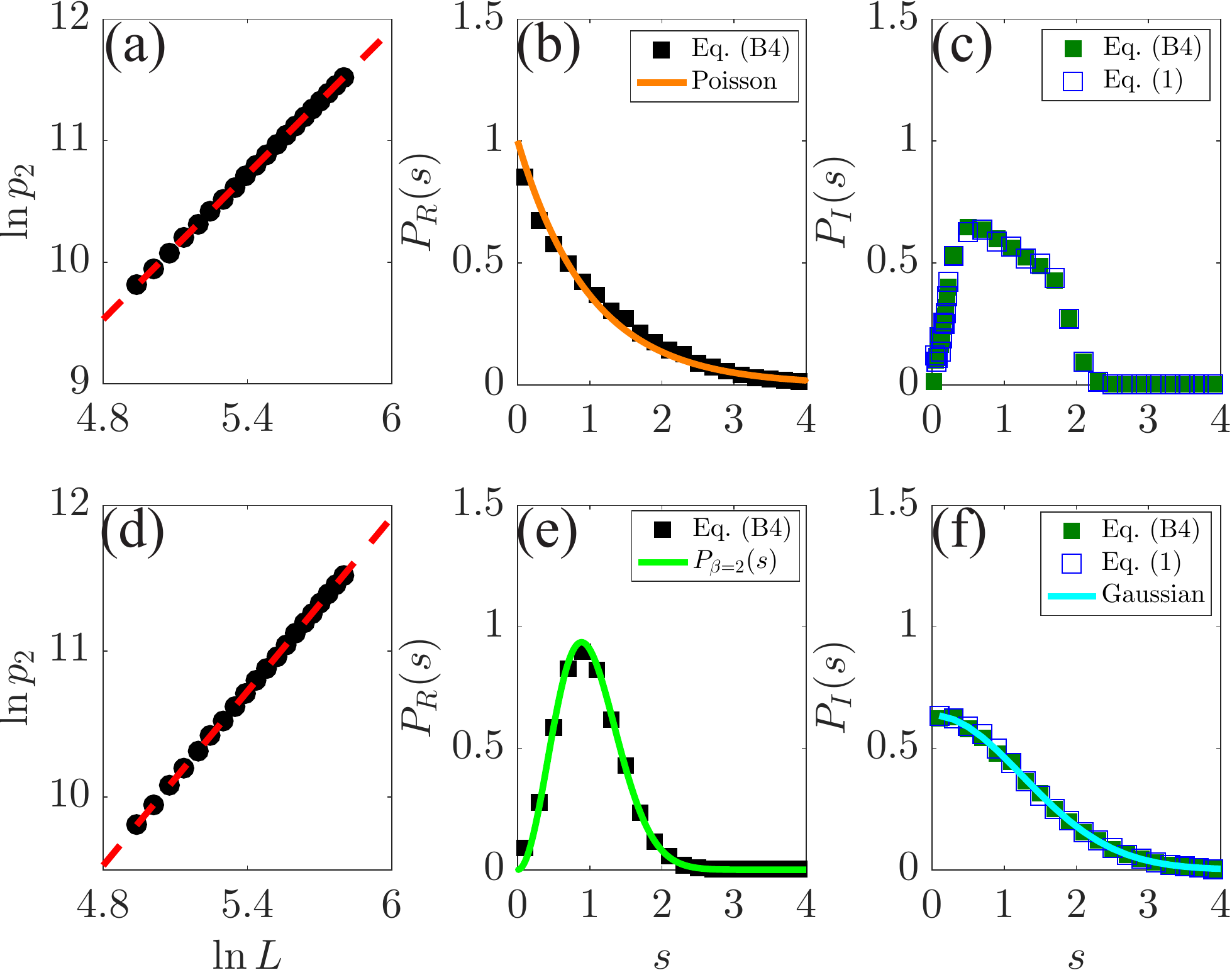}
\caption{Results of Hamiltonian~\eqref{app1_4} without TR 
symmetry ($\eta=0.1$).
(a-c) Strong Non-Hermicity $\Gamma=0.1$: 
(a) $\ln p_2(E=0)$ as a function of $\ln L$. 
The red dash line is a linear fit with slope $D=1.99\pm0.01$. 
(b) $P_R(s)$ (the black squares) within $|E|<0.01$. The orange
solid line is $P_{\text{P}}(s)$.  
(c) $P_I(s)$ within $|E|<0.01$ for Hamiltonian~\eqref{app1_4} (the 
filled squares) and for Hamiltonian~\eqref{hamiltonian} (the empty squares). 
(d-f) Weak Non-Hermicity $\Gamma=10^{-7}$: 
(d) $\ln p_2(E=0)$ as a function of $\ln L$. 
The red dash line is a linear fit with slope $D=1.99\pm0.01$.
(e) $P_R(s)$ (the black squares) within $|E|<0.01$. The green
solid line is $P_{\beta=2}(s)$. (f) $P_I(s)$ within 
$|E|<0.01$ for Hamiltonian~\eqref{app1_4} (the filled squares) and 
for Hamiltonian~\eqref{hamiltonian} (the empty squares). 
The cyan solid line is the Gaussian function.}
\label{fig5}
\end{figure}

\begin{figure}[htbp]
\centering
  \includegraphics[width=0.45\textwidth]{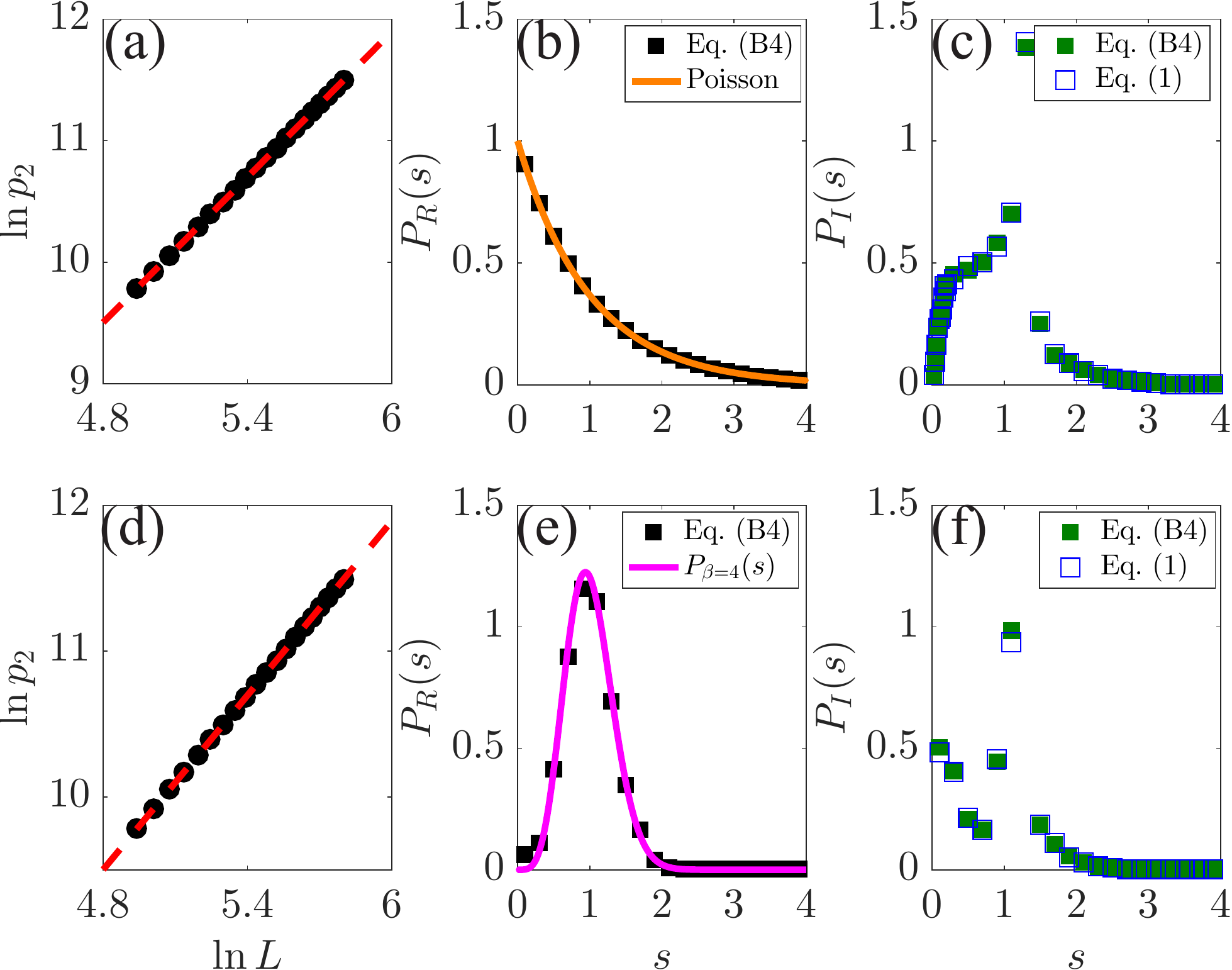}
\caption{Results of Hamiltonian~\eqref{app1_4} with TR symmetry
($\eta=0$).
(a-c) Strong Non-Hermicity $\Gamma=0.1$: 
(a) $\ln p_2(E=0)$ as a function of $\ln L$. 
The red dash line is a linear fit with slope $D=1.99\pm0.01$. 
(b) $P_R(s)$ (the black squares) within $|E|<0.01$. 
The orange solid line is $P_{\text{P}}(s)$.  
(c) $P_I(s)$ within $|E|<0.01$ for Hamiltonian~\eqref{app1_4} (the filled 
squares) and for Hamiltonian~\eqref{hamiltonian} (the empty squares). 
(d-f) Weak Non-Hermicity $\Gamma=10^{-7}$: 
(d) $\ln p_2(E=0)$ as a function of $\ln L$. 
The red dash line is a linear fit with slope $D=1.99\pm0.01$.
(e) $P_R(s)$ (the black squares) within $|E|<0.01$. 
The magenta solid line is $P_{\beta=4}(s)$. (f) $P_I(s)$ within 
$|E|<0.01$ for Hamiltonian~\eqref{app1_4} (the filled squares) and 
for Hamiltonian~\eqref{hamiltonian} (the empty squares). 
}
\label{fig6}
\end{figure}

Secondly, we show that the level statistics do not depend on 
the forms of disorders by considering the following model,
\begin{equation}
\begin{gathered}
H=\sum_{\bm{i}}c^\dagger_{\bm{i}}
(\epsilon_{\bm{i}}\sigma_0+\eta\sigma_z+i\gamma_{\bm{i}}\sigma_z)
c_{\bm{i}}+\sum_{\langle\bm{ij}\rangle}c^\dagger_{\bm{i}}V_{\bm{ij}}
c_{\bm{i}}+h.c.,
\end{gathered}\label{app1_4}
\end{equation}
where $\epsilon_{\bm{i}}$ and $\gamma_{\bm{i}}$ are 
independent random numbers that distribute in the range of 
$[-W/2,W/2]$ and $[-\Gamma/2,\Gamma/2]$, respectively. 
$V_{\bm{ij}}=V_x=\sigma_0+i\alpha\sigma_y$ 
and $V_y=\sigma_0-i\alpha\sigma_x$ for $\langle\bm{ij}\rangle$ 
along the $x-$ and the $y-$directions. 
$\alpha$ and $\eta$ are two constants measuring SOC 
strength and the degree of TR symmetry violation. Different from 
model~\eqref{hamiltonian} with the constant non-Hermicity, 
both the Hermitian and the non-Hermitian parts are random here. 
All states of this model within the energy window of 
$|E|<0.01$ for $W=1$, $\alpha=\eta=0.1$, $L=160$ 
(system sizes), and $\Gamma=0.1$ (strong non-Hermicity) 
and $10^{-7}$ (weak non-Hermicity) are extended. 
The corresponding $P_R(s)$ and $P_I(s)$ of those states 
are plotted in Figs.~\ref{fig5} ($\eta=0.1$, without 
TR symmetry) and \ref{fig6} ($\eta=0$, with TR symmetry). 
They are the same as those of Model~\eqref{app1_1}.
\par

\begin{figure}[htbp]
\centering
\includegraphics[width=0.45\textwidth]{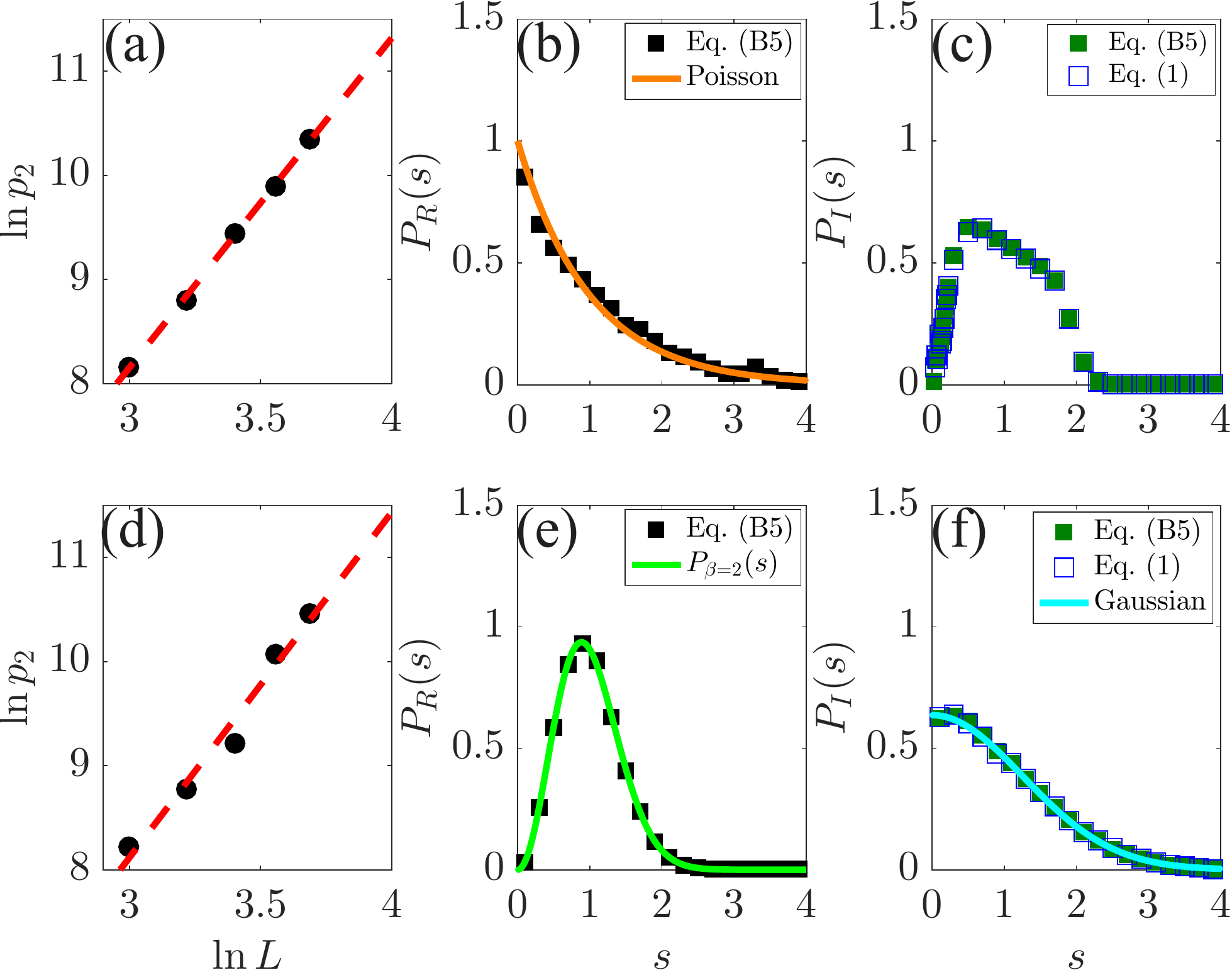}
\caption{Results of Hamiltonian~\eqref{app1_5}.
(a-c) Strong Non-Hermicity $\Theta=0.1$: (a) 
$\ln p_2(E=0)$ as a function of $\ln L$. 
The red dash line is a linear fit with slope $D=3.01\pm0.01$. 
(b) $P_R(s)$ (the black squares) within $|E|<0.01$. 
The orange solid line is $P_{\text{P}}(s)$.  
(c) $P_I(s)$ within $|E|<0.01$ for Hamiltonian~\eqref{app1_5} (the filled 
squares) and for Hamiltonian~\eqref{hamiltonian} (the empty squares). 
(d-f) Weak Non-Hermicity $\Theta=10^{-7}$: 
(d) $\ln p_2(E=0)$ as a function of $\ln L$. 
The red dash line is a linear fit with slope $D=2.98\pm0.01$.
(e) $P_R(s)$ (the black squares) within $|E|<0.01$. 
The green solid line is $P_{\beta=2}(s)$. (f) $P_I(s)$ within 
$|E|<0.01$ for Hamiltonian~\eqref{app1_5} (the filled squares) 
and for Hamiltonian~\eqref{hamiltonian} (the empty squares). 
The cyan solid line is the Gaussian function.}
\label{fig7}
\end{figure}

Thirdly, we investigate the level statistics of a three-dimensional 
non-Hermitian Anderson model 
\begin{equation}
\begin{gathered}
H=\sum_{\bm{i}}c^\dagger_{\bm{i}}(\epsilon_{\bm{i}}+i\theta_{\bm{i}})
c_{\bm{i}}+t\sum_{\langle\bm{ij}\rangle}c^\dagger_{\bm{i}}c_{\bm{j}}
+h.c.,
\end{gathered}\label{app1_5}
\end{equation}
where $c^\dagger_{\bm{i}}$ and $c_{\bm{i}}$ are the creation and 
annihilation operator of a single electron at site $\bm{i}=(l,m,n)$ 
with $l,m,n$ being integers and $1\leq l,m,n\leq L$. 
The hopping energy $t$ is chosen as the energy unit, i.e., $t=1$. 
Randomness is introduced through random real numbers $\epsilon_{\bm{i}}$ 
and $\theta_{\bm{i}}$ uniformly and independently distributed in 
$[-W/2,W/2]$ and $[-\Theta/2,\Theta/2]$, respectively. 
Periodic boundary conditions are applied in all directions to avoid the 
non-Hermitian skin effect. The obtained $P_R(s)$ and $P_I(s)$ in the energy 
interval of $|E|\in[-0.01,0.01]$ and $W=1$ are shown in Fig.~\ref{fig7}. 
Clearly, they also follow the same level statistics as those of states of 
Hamiltonian~\eqref{hamiltonian}. 
\par

\section{Mean level spacing of Hermitian part}
\label{app2}

The mean level spacing $\langle\Delta_0\rangle$ of 
the Hermitian part of Hamiltonian~\eqref{hamiltonian} is 
an important energy scale related different level statistics. 
In this section, we want to find an accurate estimate of 
$\langle\Delta_0\rangle$ 
for a given system size $L$ and disorder strength $W$.
For small disorders $W$, all eigenenergies should 
lie in the energy range of $[-(4+W/2), (4+W/2)]$ such
that the energy bandwidth is about $8+W$. Since the 
number of eigenstates is proportional to $L^2$, the 
mean level spacing should then satisfy 
\begin{equation}
\begin{gathered}
\langle \Delta_0\rangle=\beta\dfrac{(W+8)}{L^2},
\end{gathered}\label{app2_1}
\end{equation} 
with $\beta$ being a coefficient that is obtained below. 
\par

\begin{figure}[htbp]
\centering
\includegraphics[width=0.35\textwidth]{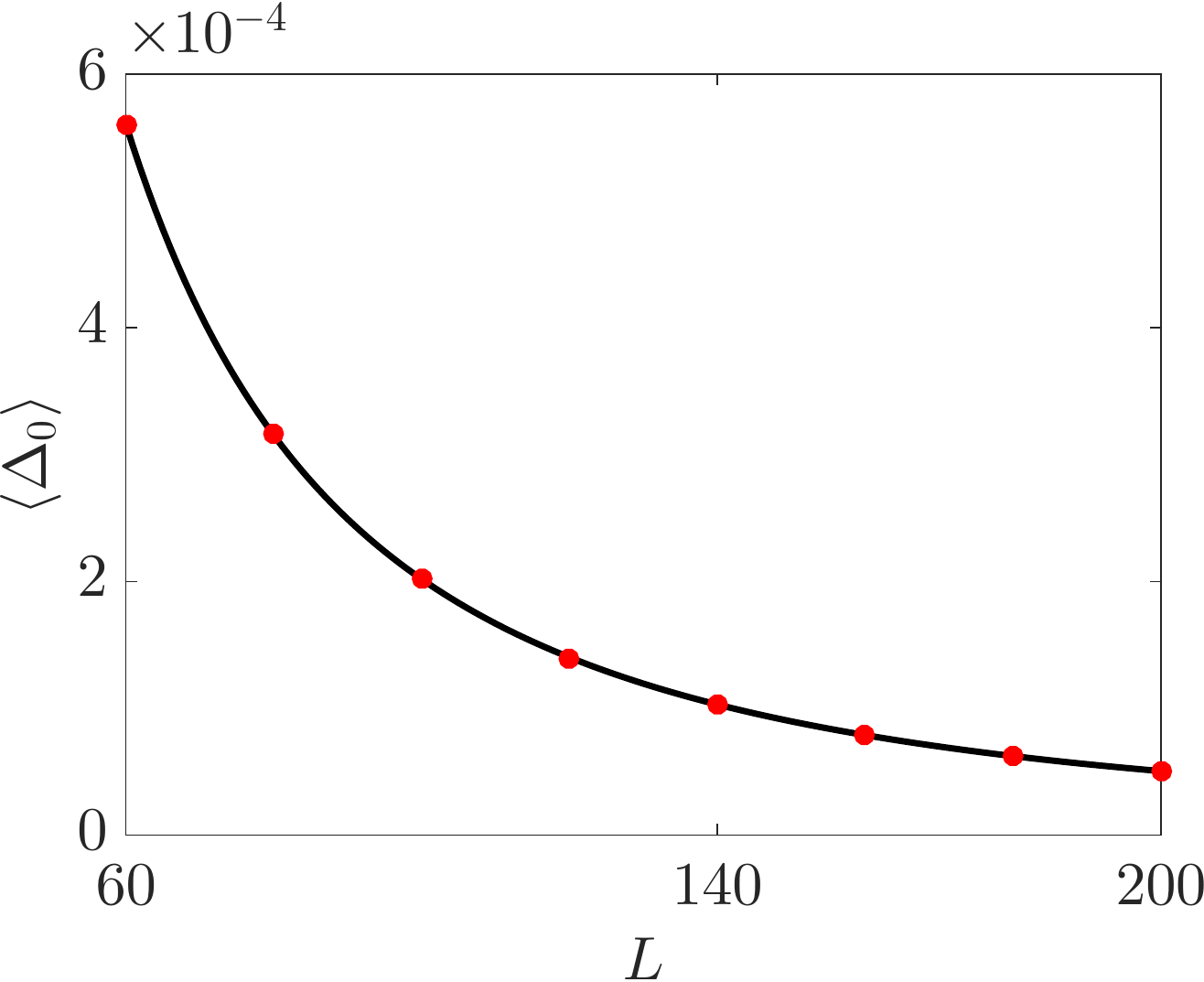}
\caption{Mean level spacings $\langle\Delta_0\rangle$ (the red 
circles) of the Hermitian part of Hamiltonian~\eqref{hamiltonian} 
as a function of $L$ for $W=1$. 
The black line is Eq.~\eqref{app2_1} with $\beta=0.22$. 
Each point is averaged over more than 200 ensembles.}
\label{fig8}
\end{figure}

To numerically determine the coefficient $\beta$, we 
calculate $\langle\Delta_0\rangle$ and plot them 
(symbols) against $L$ in Fig.~\ref{fig8}. Here 
$\Delta_0$ is obtained from a small energy window 
$[-0.01,0.01]$ around $E=0$, and $\langle\cdots\rangle$ 
is averaged over more than 200 ensembles. 
A fit of $\langle\Delta_0\rangle$ to Eq.~\eqref{app2_1} 
yields $\beta\simeq 0.22$, which accords well with numerical 
data (up to $L=200$), see the black line in Fig.~\ref{fig8}. 
Thus, the mean level spacing of the Hermitian part of 
Hamiltonian~\eqref{hamiltonian} can be obtained by formula 
$\langle\Delta_0\rangle\simeq 0.22(W+8)/L^2$.
\par

\section{Level statistics of Hamiltonian~\eqref{hamiltonian} with an additional term}
\label{app3}

We show that an additional term of $-i\gamma_0\sigma_0$ to 
Hamiltonian~\eqref{hamiltonian}, i.e., $\tilde{H}=H-\sum_{\bm{i}}
c^\dagger_{\bm{i}}\gamma_0\sigma_0 c_{\bm{i}}$, does not affect 
$P_R(s)$ and $P_I(s)$. Suppose $|\psi_E\rangle$ is an arbitrary 
right eigenstate of Hamiltonian $H$ with eigenenergy $E$, then
\begin{equation}
\begin{gathered}
\tilde{H}|\psi_E\rangle=\left(H-\sum_{\bm{i}}c^\dagger_{\bm{i}}
\gamma_0\sigma_0 c_{\bm{i}}\right)|\psi_E\rangle\\
=(H-i\gamma_0 I)|\psi_E\rangle
=(E-i\gamma_0)|\psi_E\rangle
\end{gathered}
\end{equation}
with $I$ being the identity matrix. Thus, $|\psi_E\rangle$
is also a right eigenstate of $\tilde{H}$ with eigenenergy $E-i\gamma_0$. 
Since $|\psi_E\rangle$ is arbitrary, all the levels of $\tilde{H}$ are the same 
as those of $H$ but shift by a constant imaginary value of $-i\gamma_0$. 
Obviously, the constant shift of all levels of complex energies does not 
change the distributions of level spacings, $P_R(s)$ and $P_I(s)$.
\par

\end{document}